\documentclass[10pt,conference]{IEEEtran}
\usepackage{epsfig,graphicx,subfigure,psfrag,amsmath,cases,bm}
\usepackage{latexsym,amssymb,algorithm,mathtools}
\usepackage{algorithmic}
\usepackage{color}
\usepackage{url}
\usepackage{scrtime}
\usepackage{stfloats}
\usepackage{tablefootnote}
\usepackage{cite}
\usepackage{amsfonts,mathabx}
\usepackage{textcomp}
\usepackage{xcolor,capt-of}
\usepackage{multirow}

\newcommand*{\rom}[1]{\expandafter\@slowromancap\romannumeral #1@}
\makeatother

\usepackage[left=0.625in,top=0.7in,bottom=0.99in,right=0.625in]{geometry}
\author{\IEEEauthorblockN {Ata Khalili, Brikena Kaziu, Vasilis K. Papanikolaou, and Robert Schober\\
Friedrich-Alexander-University
Erlangen-N\"urnberg, Germany;
}

\thanks{This work was supported partly by the Federal Ministry of Education and Research of Germany under the program of “Souveran. Digital. Vernetzt.” joint project 6G-RIC (project identification number: PIN 16KISK023) and also in part by the Deutsche Forschungsgemeinschaft (DFG, German Research Foundation) GRK-2680 – Project-ID 437847244.}}

\newtheorem{T-Prob}{Transformed Problem}

\allowdisplaybreaks

\title{\huge Pinching Antenna-enabled ISAC Systems: Exploiting Look-Angle Dependence of RCS for Target Diversity}

\begin{document}
\maketitle
\begin{abstract}
We investigate a novel integrated sensing and communication (ISAC) system supported by pinching antennas (PAs), which can be dynamically activated along a dielectric waveguide to collect spatially diverse observations. This capability allows different PAs to view the same target from different angles across time, thereby introducing target diversity, which is a key advantage over conventional fixed antenna arrays. To quantify the sensing reliability, we adopt the outage probability as a performance metric, capturing the likelihood that the accumulated radar echo signal power falls below a detection threshold. In contrast to traditional ISAC models that assume deterministic sensing channels, we explicitly account for the look-angle dependence of radar cross-section (RCS) by modeling it as a random variable. We ensure the long-term quality-of-service (QoS) for communication users by enforcing an accumulated data rate constraint over time. We derive an exact closed-form expression for the sensing outage probability based on the distribution of weighted sums of exponentially distributed random variables. Since the resulting expression is highly non-convex and intractable for optimization, we use a tractable upper bound based on the Chernoff inequality and formulate a PA activation optimization problem. A successive convex approximation (SCA) framework is proposed to efficiently solve the formulated problem. Numerical results show that dynamically activating different PAs across time slots significantly enhances sensing reliability compared to repeatedly activating the same PA at a fixed position and conventional antenna selection schemes, respectively. These findings highlight the benefits of integrating outage-based reliability metrics and target diversity into ISAC systems using PAs.
\end{abstract}
\section {Introduction}
The sixth-generation (6G) wireless networks are envisioned to support both ultra-reliable, high-speed communication, and real-time environmental awareness. To meet these dual demands, integrated sensing and communication (ISAC) has emerged as a key enabler for future wireless systems, allowing the same infrastructure and spectrum to be used for both communication and radar sensing tasks. Multiple-input multiple-output (MIMO) technologies have played a central role in enabling ISAC by exploiting spatial diversity to improve both data throughput and sensing resolution. However, conventional MIMO architectures rely on fixed antenna arrays, which limits spatial adaptability, particularly in dynamic or harsh environments where multipath effects and signal blockages can impair both sensing and communication performance. To address these challenges, advanced flexible-antenna systems, such as fluid antennas and movable antennas, have gained attention for their ability to dynamically reposition antenna elements, providing greater adaptability to channel variations and enhancing spatial diversity\cite{wong2020fluid,zhu2023movable,khalili2024advanced}.

Recently, pinching antennas (PAs) have been proposed as a novel flexible-antenna technology that addresses the limitations of conventional antenna systems \cite{docomo2019ntt,p1}. Unlike other flexible-antenna designs, PAs utilize dielectric waveguides, allowing antennas to be dynamically activated at any point along the waveguide. This design enables flexible and scalable deployment, as antennas can be positioned closer to users to establish strong line-of-sight (LoS) links. Unlike existing systems, which are often constrained within apertures spanning a few wavelengths, dielectric waveguides can be significantly longer, making it possible to mitigate the negative impact of large-scale path loss. Additionally, PAs are cost-effective, easy to deploy, and offer high adaptability, as the waveguide structure allows for dynamic addition and removal of antennas without significant overhead. Previous works have demonstrated the performance gains enabled by PAs in multiuser and uplink communication scenarios, highlighting their potential to outperform conventional fixed-antenna systems \cite{p3,p4}.

While ISAC promises efficient spectrum and hardware reuse, a major challenge for reliable sensing lies in the angular dependence of the radar cross-section (RCS). In the radar literature, it is well established that the RCS depends on the target’s geometry, material properties, and look-angle, and its value is typically modeled as a Gaussian random variable\cite{Radar}. However, this effect is largely overlooked in existing ISAC literature \cite{ISAC6G,jsc-mimo-radar,mu-mimo-jsc}. For static targets, the RCS remains constant over time for a given look-angle but varies significantly with the look-angle due to the target’s geometry and material properties. As a result, systems that observe the target from a single angle may suffer from low sensing performance if the corresponding RCS is small. To address this issue, distributed MIMO radar systems employing spatially widely separated antennas have been proposed to achieve \textit{target diversity} by observing the same target from different angles~\cite{Ang1,Ang2}. Inspired by this idea, this work proposes the exploitation of \textit{target diversity} within an ISAC system by dynamically activating PAs at different positions along a waveguide, thereby enabling multiple angular perspectives. Conventional multi-antenna systems used in ISAC, including fixed antenna arrays and flexible architectures like movable or fluid antennas, cannot provide target diversity. These systems are confined to small transmit areas and thus are not capable of exploiting angular diversity and mitigating the effects of unfavorable RCS look-angles. 

 To systematically address the challenges introduced by the look-angle dependence of the RCS, we propose the \textit{radar outage probability} as a practical and rigorous performance metric for sensing in ISAC systems. This metric reflects the likelihood that the accumulated radar signal-to-noise ratio (SNR) across multiple angular views falls below a given detection threshold, thus quantifying the reliability of the sensing operation under RCS uncertainty. While outage probability is a well-established concept in communication theory, its application to radar sensing is new, particularly within ISAC frameworks. We formulate a tractable optimization framework that activates PAs at different positions over multiple time slots to minimize the outage probability, thereby fully exploiting the angular diversity enabled by PAs. Although the potential of PAs for ISAC was recently acknowledged in \cite{P_ISAC}, the concrete benefits of target diversity and the integration of outage-based reliability modeling for system optimization have not been explored, yet. This paper fills this gap by establishing both theoretical and algorithmic foundations for robust and spatially adaptive PA-assisted ISAC design. The main contributions of this paper are summarized as follows:
\begin{itemize}
    \item We investigate, for the first time, the integration of PAs into ISAC systems and highlight their ability to achieve \emph{target diversity} by providing multiple angular views of the same target across time.
    \item We introduce \emph{radar outage probability} as a novel performance metric for radar sensing in ISAC, capturing the loss in reliability due to the look-angle dependence of the RCS. A closed-form expression for the outage probability is derived using weighted sums of exponentially distributed random variables.
    \item To enable tractable optimization, we propose a Chernoff bound-based approximation of the outage probability and formulate a PA activation problem.
    \item Our results show that dynamic and spatially distributed PA activation significantly improves sensing reliability, validating the benefits of PAs and outage-based design of ISAC systems.
\end{itemize}

\textit{Notation:} 
In this paper, matrices and vectors are denoted by
boldface capital letters $\mathbf{A}$ and lower case letters $\mathbf{a}$, respectively.~$\mathbf{A}^T$, $\mathbf{A}^{*}$, and $\mathbf{A}^H$ are the transpose,~conjugate, and Hermitian of matrix $\mathbf{A}$, respectively. $|\cdot|$ and $||\cdot||_2$ stand for the absolute value of a complex scalar and the $l_2$-norm of a vector, respectively. $\mathbf{I}_N$ is the $N$-by-$N$ identity matrix. $\mathbb{R}^{N\times M}$ and $\mathbb{C}^{N\times M}$ represent the space of $N\times M$ real-valued and complex-valued matrices, respectively. $\mathrm{diag}(\mathbf{a})$ denotes a diagonal matrix whose main diagonal elements are given by the entries of vector $\mathbf{a}$. $\mathbb{E}[\cdot]$ refers to statistical expectation.
\begin{figure}
    \centering
    \includegraphics[trim=0 2.2cm 13cm 0, clip, width=0.82\linewidth]{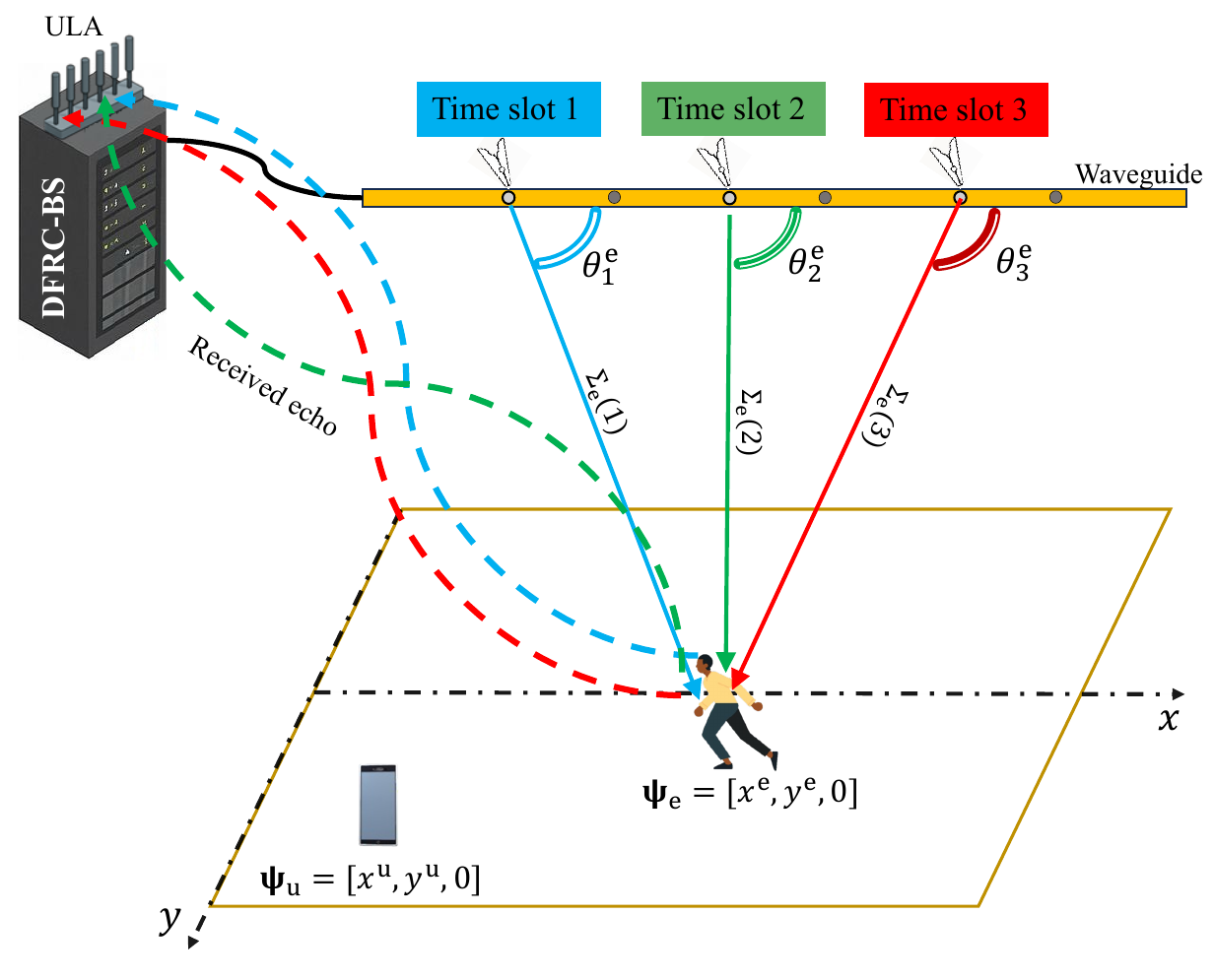}
    \caption{An illustration of the considered PA-assisted ISAC system with a DFRC-BS transmitter, dynamically activated PAs, a communication user, a radar receiver, and a target. By activating PAs at different positions across time slots, the system achieves multiple angular views, enabling target diversity. $\Sigma_{e}(t)$ denotes the RCS coefficient in time slot $t$, which depends on the corresponding look-angle $\theta^{e}_{m}$.}
    \label{fig:PA_system_model}
\end{figure}
\section{System Model}
We consider an ISAC system employing a dual-function radar and communication (DFRC) base station (BS) equipped with a set of $M$ preconfigured PA positions along a dielectric waveguide, see Fig. \ref{fig:PA_system_model}. While there are $M$ candidate positions available, only one PA element is activated per time slot. Time is divided into \(T\) time slots, indexed by \(t \in \mathcal{T} = \{1, 2, \dots, T\}\). In each time slot, the DFRC-BS simultaneously serves one single-antenna communication user and performs radar sensing of a target. The dynamic activation of the PAs enables the BS to observe the target from different angular directions $\theta^{e}_{m}$, $\forall m \in \mathcal{M}=\{1,...,M\}$, determined by the active PA position in each time slot. This introduces \textit{target diversity}, which helps mitigate the look-angle dependence of the RCS and enhances sensing reliability. To be able to suppress self-interference and improve sensing performance, we assume that the radar echoes are collected at the BS by a separate fixed receive uniform linear array (ULA) equipped with \(N_R\) antennas. This hybrid configuration combines the spatial agility of PAs for transmission, enabling time-varying directional probing of the target, with the stability of a fixed receive array, which ensures reliable signal capture and allows phase-aligned processing of the radar echoes for enhanced detection and estimation accuracy.

\subsection{Pinching Antenna Configuration}
The dielectric waveguide is installed parallel to the $x$-axis at a fixed height $d$. It hosts a set of preconfigured discrete positions, where PAs can be dynamically activated. Let $\mathcal{P} = \{\mathbf{p}_{1}, \mathbf{p}_{2}, \dots, \mathbf{p}_{M}\}$ denote the set of $M$ preconfigured feasible PA positions along the waveguide, where the $m$-th position is defined by $\mathbf{p}_{m} = (x_{m}, 0, d)$, with $x_{m} \in [0,D_{x}]$ denoting the horizontal position and $d$ the fixed vertical placement above the plane. In each time slot \( t \in \mathcal{T} \), exactly one PA is activated at the selected position from the predefined set \( \mathcal{P} \). We define a binary selection variable \( b_m(t) \in \{0,1\} \), such that:
\begin{equation}
    b_m(t) = 
    \begin{cases}
    1, \, \text{if the PA at } \mathbf{p}_m~\text{is activated }\text{in time slot } t, \\
    0, \, \text{otherwise}.
    \end{cases}
\end{equation}

To ensure only one PA is activated at a time, the selection must satisfy the constraint:
\begin{equation}
    \sum_{m=1}^{M} b_{m}(t) = 1, \quad \forall t \in \mathcal{T}.
\end{equation}

We collect the binary position selection variables in  vector $\mathbf{b}(t) = [b_1(t), b_2(t), \dots, b_M(t)]^T \in \{0,1\}^{M}$. The set of predefined positions $\mathcal{P}$ is designed with sufficient spatial separation between elements to ensure distinct angular views of the target. Although only one PA is active at any given time, this spacing is critical to ensure that each activated position corresponds to a different RCS providing meaningful angular diversity for sensing.

\subsection{Signal Model}

In each time slot \( t \in \mathcal{T} \), the DFRC-BS transmits a signal \( x(t) \in \mathbb{C} \) via a single active PA to simultaneously serve one communication user and perform radar sensing. The signal carries an information symbol \( c(t) \in \mathbb{C} \), intended for communication, which is also used for radar probing. The transmit signal is given by
\begin{equation}
    x(t) = \sqrt{p_t}\, c(t),
\end{equation}
where \( p_t \) denotes the transmit power and \( c(t) \) is a unit-power information-bearing symbol with \( \mathbb{E}[|c(t)|^2] = 1 \).
\subsection{Communication Channel and Metric}

In the proposed PA-enabled ISAC system, the communication channel is strongly influenced by the relative positions of the communication user and the dynamically selected PA location along the waveguide. The channel between BS and user for each candidate PA position includes two key components: (i) the free-space path loss and phase shift introduced by the signal propagation from the PA to the user, and (ii) the attenuation and phase shift incurred by the waveguide. The composite channel vector of the user for all PA positions, $\hat{\mathbf{h}}_u$ is specified in (4) at the top of this page.
\begin{figure*}[t]
\begin{align}
\hat{\mathbf{h}}_{u} = \left[
\frac{\eta e^{-j \frac{2\pi}{\lambda} \|\boldsymbol{\psi}_u - \mathbf{p}_{1}\|}}{\|\boldsymbol{\psi}_u - \mathbf{p}_{1}\|}e^{(-\alpha - j \frac{2\pi}{\lambda_g}) \|\mathbf{p}_{0} - \mathbf{p}_{1}\|}, \, \cdots, \,
\frac{\eta e^{-j \frac{2\pi}{\lambda} \|\boldsymbol{\psi}_u - \mathbf{p}_{M}\|}}{\|\boldsymbol{\psi}_u - \mathbf{p}_{M}\|}e^{(-\alpha - j \frac{2\pi}{\lambda_g}) \|\mathbf{p}_{0} - \mathbf{p}_{M}\|}
\right]^T
\end{align}
\vspace{-5mm}
\end{figure*}
Here, \( \boldsymbol{\psi}_u \) denotes the location of the user, and \( \eta = \frac{c}{4\pi f_c} \) is the free-space path loss constant, with \( c \) representing the speed of light and \( f_c \) the carrier frequency. The second exponential term in each element of $\hat{\mathbf{h}}_u$ captures both the amplitude attenuation and phase shift introduced by wave propagation through the dielectric waveguide, where \( \lambda_g = \lambda / n_{\mathrm{eff}} \) denotes the guided wavelength determined by the effective refractive index $n_{\mathrm{eff}}$, and $\alpha$ represents the waveguide's attenuation coefficient. \( \mathbf{p}_0 \) denotes the feed point of the waveguide. The effective channel gain of the user in time slot \( t \) is given by $h(t) = \hat{\mathbf{h}}_{u}^H \mathbf{b}(t)$.

The received baseband signal at the communication user is given by:
\begin{equation}
    y(t) = h(t) x(t) + z(t),
\end{equation}
where \( z(t) \sim \mathcal{CN}(0, \sigma^2) \) denotes additive white Gaussian noise. Accordingly, the instantaneous SNR at the communication user in time slot \( t \) is
\begin{equation}
    \gamma(t) = \frac{p_t}{\sigma^2} \left| h(t) \right|^2.
\end{equation}

\subsection{Sensing Channel and Metric}
For the target, the channel gain corresponding to all \( M \) candidate PA positions along the waveguide that captures the impact of both the free-space and waveguide propagation from the feed point \( \mathbf{p}_0 \) to the target via the PAs is given in \eqref{channel_sens} at the top of this page, where \( \boldsymbol{\psi}_e \) denotes the target’s position.
\begin{figure*}
    \begin{align}\label{channel_sens}
\hat{\mathbf{h}}_{e} = \big[ &\frac{e^{-j \frac{2\pi}{\lambda} \|\boldsymbol{\psi}_e - \mathbf{p}_1\|}}{\|\boldsymbol{\psi}_e - \mathbf{p}_1\|} e^{(-\alpha - j \frac{2\pi}{\lambda_g})\|\mathbf{p}_0 - \mathbf{p}_1\|}, \, \dots,\frac{e^{-j \frac{2\pi}{\lambda} \|\boldsymbol{\psi}_e - \mathbf{p}_M\|}}{\|\boldsymbol{\psi}_e - \mathbf{p}_M\|} e^{(-\alpha - j \frac{2\pi}{\lambda_g})\|\mathbf{p}_0 - \mathbf{p}_M\|} \big]^T
\end{align}
\hrule
\end{figure*}

\subsubsection{Radar Echo Model}
 Assuming a narrowband LoS channel and a stationary target \cite{jsc-mimo-radar}, the received radar echo at the BS in time slot \( t \) is modeled as:
\begin{equation}
    \mathbf{r}_e(t) = \mathbf{g}(t) x(t) + \mathbf{z}(t),
\end{equation}
where \( \mathbf{g}(t) \in \mathbb{C}^{N_R \times 1} \) is the round-trip sensing channel, and \( \mathbf{z}(t) \sim \mathcal{CN}(\mathbf{0}, \sigma^2 \mathbf{I}_{N_R}) \) is additive white Gaussian noise. The sensing channel vector in time slot $t$ is given by
\begin{equation}
\mathbf{g}(t) = \frac{\beta_0 \Sigma_e(t)}{d_{e,r}} \cdot \mathbf{a}_r(\theta_e)  \hat{\mathbf{h}}^{H}_{e}\mathbf{b}(t),
\end{equation}
where $\beta_0$ denotes the large-scale path loss at a reference distance $d_0 = 1$ meter, and $d_{e,r}$ is the distance from the target to the receiver array. Vector $\mathbf{a}_r(\theta_e) \in \mathbb{C}^{N_R \times 1}$ is the receive array steering vector corresponding to the target’s direction\cite{khalili2024efficient}. The RCS coefficient \( \Sigma_e(t) \) is modeled as a zero-mean circularly symmetric complex Gaussian random variable, i.e.,  \( \Sigma_e(t) \sim \mathcal{CN}(0, \Sigma_{\mathrm{av}}) \). While the target is static, the reflected signal strength depends on the direction from which it is observed a phenomenon known as angular dependence of the RCS\cite{Ang1}. In our setup, a different PA is activated in each time slot, and each PA position corresponds to a distinct look-angle towards the target. Because the RCS varies with the look-angle, each activated PA leads to a different realization of the RCS. Therefore, we model \( \Sigma_e(t)\) as an independent and identically distributed (i.i.d.) complex Gaussian random variable \cite{Radar}, where independence across time slots is justified by ensuring sufficient spatial separation between the preconfigured PA positions and the resulting uncorrelated scattering responses from different observation angles.
 After applying receive beamforming vector \( \mathbf{u} \in \mathbb{C}^{N_{R}\times 1} \), the received sensing signal becomes:
\begin{equation}
    \tilde{r}(t) = \mathbf{u}^H \mathbf{g}(t) x(t) + \mathbf{u}^H \mathbf{z}(t).
\end{equation}

We adopt receive beamforming vector \( \mathbf{u} = \frac{\mathbf{a}_r(\theta_e)}{\|\mathbf{a}_r(\theta_e)\|} \), aligned with the known steering direction of the target. Applying this beamformer to the received signal, the instantaneous radar SNR in time slot $t$ is given by:
\begin{equation}\label{SNR}
    \Gamma(t) = \frac{p_t}{\sigma^2} \left| \mathbf{u}^H \mathbf{g}(t) \right|^2.
\end{equation}
Substituting the expression for \( \mathbf{g}(t) \) into the above equation and simplifying yields:
\begin{equation}
    \Gamma(t) = \psi(t) |\Sigma_e(t)|^2,
\end{equation}
where the deterministic gain factor is given by:
\begin{equation}
    \psi(t) = \frac{p_t \beta_0^2}{ \sigma^2 d_{e,r}^2}  \|\mathbf{a}_r(\theta_e)\|^2  \left| \hat{\mathbf{h}}_{e}^{H}\mathbf{b}(t)\right|^2.
\end{equation}

\subsubsection{Radar Outage Probability}
To evaluate the reliability of radar sensing, we define the \textit{radar outage probability} as the likelihood that the accumulated SNR over all time slots falls below a predefined threshold \( \Gamma^{\mathrm{th}} \). In our setup, a single PA is activated in each time slot, and the corresponding radar echo depends on the look-angle toward the target. Since the RCS varies with angle, each echo corresponds to a distinct and statistically independent realization. This motivates the use of outage probability as a meaningful metric to capture the impact of variations in the radar return signal across time slots. Over \( T \) time slots, the accumulated sensing SNR is given by:
\begin{equation} \label{eq:targetSNR}
    \Gamma_{\text{total}} = \sum_{t=1}^{T} \Gamma(t) = \sum_{t=1}^{T} \psi(t) |\Sigma_e(t)|^2,
\end{equation}  
where \( \psi(t) \) is the deterministic gain in time slot \( t \). The radar outage probability is then defined as: \\ \\
\begin{equation} \label{eq:outage1}
    P_{\mathrm{out}} = \Pr\left( \Gamma_{\text{total}} < \Gamma^{\mathrm{th}} \right).
\end{equation}
This corresponds to the cumulative distribution function (CDF) of a weighted sum of independent, non-identically exponentially distributed random variables, $ \psi(t)|\Sigma_e(t)|^2$. 
\section{Problem Formulation}
We consider the minimization of the radar outage probability as the primary  objective for sensing. This metric captures the probability that the accumulated sensing SNR over a transmission period falls below a predefined threshold, resulting in unreliable target detection. The choice of outage probability is motivated by its practical relevance in dynamic environments, where RCS variations introduce stochastic uncertainty regarding the received echo strength.
We formulate a PA activation optimization problem, where our goal is to determine the PA selection strategy that minimizes the sensing outage probability while ensuring a minimum quality-of-service (QoS) for the communication user. The resulting problem is formulated as follows:
\begin{align}
\label{Ori_Problem_1}
\mathcal{P}_{0}: \quad 
&\underset{\{\mathbf{b}(t)\}_{t=1}^{T}}{\text{minimize}} \quad \Pr\left( \sum_{t=1}^{T} \psi(t) |\Sigma_e(t)|^2 < \Gamma^{\text{th}} \right)\nonumber \\
\text{s.t.} \quad
&\text{C1: } \sum_{t=1}^{T}\log_2(1+\gamma(t)) \geq R_{\min}, \quad \nonumber \\
&\text{C2: } \sum_{m=1}^{M} b_m(t) = 1, \quad \forall t \in \mathcal{T}, \nonumber \\
&\text{C3: } \sum_{t=1}^{T} b_m(t) \leq 1, \quad \forall m \in \mathcal{M}, \nonumber \\
&\text{C4: } b_m(t) \in \{0,1\}, \quad \forall m \in \mathcal{M}, \forall t \in \mathcal{T}, 
\end{align}
where C1 ensures that the user’s accumulated communication rate over the transmission period exceeds the minimum required rate \( R_{\min} \). C2 guarantees that exactly one PA is activated per time slot. C3 ensures that each PA position is used in at most one time slot to avoid repeated observations from the same angle. This promotes angular diversity and ensures that the RCS realizations across time slots remain statistically independent. C4 indicates that the position selection indicator for the PAs is a binary variable, i.e., each antenna is either activated or deactivated in each time slot. Problem \( \mathcal{P}_0 \) is inherently non-convex due to the probabilistic objective function and the binary nature of the decision variables. In the next section, we present a tractable approximation and optimization strategy based on convex relaxation and bounding techniques.
\section{Proposed Solution}
We first derive a closed-form expression for the radar outage probability. The radar outage probability in \eqref{eq:outage1} depends on the RCS \( |\Sigma_e(t)|^2 \), which is assumed to follow an exponential distribution with mean \( \Omega_{\mathrm{av}} \), i.e., \( |\Sigma_e(t)|^2 \sim \text{Exp}(1/\Omega_{\mathrm{av}}) \)\cite{khalili2024advanced,Radar}. This reflects the statistical variation of the RCS over time due to changes in the target’s angular appearance. 
Since $\Gamma_{\text{total}}$ in \eqref{eq:targetSNR} is a weighted sum of independent, non-identically exponentially distributed random variables \cite{akkouchi2008convolution}, the cumulative distribution function (CDF) of \( \Gamma_{\text{total}} \) can be written in closed form as: \\ \\ \\
\begin{equation}
F_{\Gamma_{\mathrm{total}}}(\Gamma_{\mathrm{th}}) = 1 - \sum_{t=1}^T \left( \frac{ \prod_{m=1}^T \lambda_m }{ \prod_{\substack{m=1 \\ m \neq t}}^T (\lambda_m - \lambda_t) } \cdot \frac{e^{-\lambda_t \Gamma^{\mathrm{th}} }}{\lambda_t} \right),
\label{eq:outage_cdf}
\end{equation}\\
where \( \lambda_t = \frac{1}{\psi(t)\Omega_{\mathrm{av}}} \)\footnote{The closed-form expression in \eqref{eq:outage_cdf} is valid if all \( \lambda_t \) are distinct, i.e., \( \lambda_m \neq \lambda_t \) for all \( m \neq t \). This condition is satisfied in our setup since the gain profile \( \psi(t) \) varies across time slots due to the different PA positions. For the more general case, where some $\lambda_{t}$ may be identical, the more general expression provided in \cite[Eq. (4.3)]  {akkouchi2008convolution} can be used.}. Therefore, the exact outage probability can be evaluated as $P_{\mathrm{out}} = F_{\Gamma_{\mathrm{total}}}(\Gamma_{\mathrm{th}})$.

Although the expression in \eqref{eq:outage_cdf} is useful for benchmarking, it is not suitable for optimization. In particular, $F_{\Gamma_{\mathrm{total}}}(\Gamma_{\mathrm{th}})$ is a highly nonlinear and non-convex rational function of \( \psi(t) \), which itself depends on the antenna selection vector \( \mathbf{b}(t) \). As a result, directly integrating this expression into an optimization problem is analytically intractable and numerically unstable. To overcome this difficulty, we apply the \textit{Chernoff bound}, which yields a tractable upper bound on the outage probability:
\begin{equation}
    P_{\mathrm{out}} \leq \min_{s > 0} \left\{ e^{-s \Gamma^{\mathrm{th}}} \prod_{t=1}^{T} \mathbb{E} \left[ e^{s \psi(t) |\Sigma_e(t)|^2} \right] \right\}.
\label{eq:chernoff_bound}
\end{equation}

Since $ \psi(t) |\Sigma_e(t)|^2$ follows an exponential distribution, its moment generating function is given by:
\begin{equation}
    \mathbb{E} \left[ e^{s \psi(t) |\Sigma_e(t)|^2} \right] = \frac{1}{1 - s \psi(t) \Omega_{\mathrm{av}}}, \quad \text{for } s \psi(t) \Omega_{\mathrm{av}} < 1.
\end{equation}

Substituting this into \eqref{eq:chernoff_bound} yields:
\begin{equation}
    P_{\mathrm{out}} \leq \min_{s > 0} \left\{ e^{-s \Gamma^{\mathrm{th}}} \prod_{t=1}^{T} \frac{1}{1 - s \psi(t) \Omega_{\mathrm{av}}} \right\}.
\end{equation}

To facilitate optimization, we take the logarithm of the upper bound which preserves monotonicity and yields:
\begin{equation}
    \min_{s > 0} \left\{ -s \Gamma^{\mathrm{th}} - \sum_{t=1}^{T} \log\big(1 - s \psi(t) \Omega_{\mathrm{av}} \big) \right\},
\label{eq:chernoff_objective}
\end{equation}
subject to \( s \psi(t) \Omega_{\mathrm{av}} < 1 \), $\forall t$. The objective function in \eqref{eq:chernoff_objective} is unimodal with respect to \( s \), and the optimal value can be efficiently computed via a line search over the feasible interval \( \left(0, \frac{1}{\max_t \psi(t) \Omega_{\mathrm{av}}} \right) \). This Chernoff-based formulation provides a smooth, differentiable surrogate for the outage probability, allowing us to incorporate it into a convex optimization framework.

Next, we relax the binary constraint C4 to a continuous one as $\overline{\text{C4}}$$:0\leq b_m(t) \leq 1$. Furthermore, the data rate expression for the communication user in time slot \( t \) in C1 is non-convex due to the composition of the logarithmic function with a quadratic form in \( \mathbf{b}(t) \). To address this, we apply successive convex approximation (SCA). In each SCA iteration $i$, the rate function $R(t)$ is linearized around the current solution $\mathbf{b}^{(i)}(t)$ using a first-order Taylor expansion:
\begin{equation}
    R(t)\approx \tilde{R}(t)=R^{(i)}(t) + \big(\nabla_{\mathbf{b}(t)} R(t)\big|_{\mathbf{b}^{(i)}(t)}\big)^{T} \left( \mathbf{b}(t) - \mathbf{b}^{(i)}(t) \right),
\end{equation}
where \( R^{(i)}(t) \) is the rate evaluated at \( \mathbf{b}^{(i)}(t) \), and the gradient is computed as
    $\nabla_{\mathbf{b}(t)} R(t) = \frac{p_t}{ \sigma^2} \cdot \frac{ \left( \hat{\mathbf{h}}^H \mathbf{b}^{(i)}(t) \right)^* \hat{\mathbf{h}} }{1 + \frac{p_t}{\sigma^2} \left| \hat{\mathbf{h}}^H \mathbf{b}^{(i)}(t) \right|^2 }$.
To encourage the relaxed position selection variables \( b_m(t) \in [0,1] \) to take binary values, we incorporate a penalty function $
  f(\mathbf{b}) \triangleq \sum_{t=1}^T \sum_{m=1}^M b_m(t)\left(1 - b_m(t)\right)$ to penalize non-binary solutions \cite{khalili2024advanced}, where $\mathbf{b}= [\mathbf{b}^{T}(1), \mathbf{b}^{T}(2), \dots, \mathbf{b}^{T}(T)]^T$ is the vector collecting all position selection variables across time slots. However, \( f(\mathbf{b}) \) is not convex. To address this, we linearize \( f(\mathbf{b}) \) using a first-order Taylor approximation around a reference point \( b_m^{(i)}(t) \), yielding the surrogate expression $\widetilde{f}(\mathbf{b}) = \sum_{t=1}^T \sum_{m=1}^M  b_m^{(i)}(t)(1 - b_m^{(i)}(t)) + (1 - 2 b_m^{(i)}(t))(b_m(t) - b_m^{(i)}(t))$. The linearized penalty function \( \widetilde{f}(\mathbf{b}) \) is then incorporated into the objective function with penalty weight \( \rho > 0 \). The final convex subproblem, for fixed Chernoff parameter \( s \), is formulated as follows:
\vspace{-3mm}
\begin{align}
\label{Convex_Problem}
\mathcal{P}_1: \quad &\underset{\{ \mathbf{b}(t)\}_{t=1}^{T}}{\text{minimize}} \quad 
 -s \Gamma^{\mathrm{th}} - \sum_{t=1}^{T} \log\left(1 - s \psi(t) \Omega_{\text{av}} \right) + \rho \widetilde{f}(\mathbf{b})\nonumber\\
\text{s.t.} \quad 
& \overline{\text{C1}}: \sum_{t=1}^T \tilde{R}(t) \geq R_{\min}, \nonumber \\
& \text{C2}: \sum_{m=1}^M b_m(t) = 1, \quad \forall t \in \mathcal{T}, \nonumber \\
& \text{C3}: \sum_{t=1}^{T} b_m(t) \leq 1, \quad \forall m \in \mathcal{M}, \nonumber \\
& \overline{\text{C4}}: 0 \leq b_m(t) \leq 1, \quad \forall m \in \mathcal{M},\, \forall t \in \mathcal{T}, \nonumber \\
& \text{C5}: s \psi(t) \Omega_{\text{av}} < 1, \quad \forall t \in \mathcal{T}. 
\end{align}
\noindent For a fixed value of \( s \) and in iteration $i$, problem \eqref{Convex_Problem} is convex and can be optimally solved using CVX. The SCA framework iteratively updates the linearization point $\mathbf{b}^{(i)}=[\big(\mathbf{b}^{(i)}(1)\big)^{T}, \dots, \big(\mathbf{b}^{(i)}(T)\big)^{T}]^T$ until convergence. The proposed solution is summarized in \textbf{Algorithm}~\ref{alg:sca_outage}.
\vspace{1mm}
\begin{algorithm}[t]
\caption{SCA-Based Outage Minimization with Chernoff Bound}
\label{alg:sca_outage}
\begin{enumerate}
    \item \textbf{Input:} Initial position vector $\mathbf{b}^{(0)} = \{ \mathbf{b}^{(0)}(t) \}_{t=1}^T$, Chernoff search grid over $s \in \left(0, \frac{1}{\max_t \psi(t)\Omega_{\text{av}}}\right)$, penalty weight $\rho$, tolerance $\epsilon$
    
    \item \textbf{Initialize:} Best objective value $\mathcal{F}_{\text{best}} \gets \infty$, optimal Chernoff parameter $s^* \gets 0$, optimal position vector $\mathbf{b}^* \gets \mathbf{b}^{(0)}$

    \item \textbf{For each} $s$ in the search grid:
    \begin{enumerate}
        \item Set iteration index $i=0$
        
        \item \textbf{Repeat:}
        \begin{itemize}
            \item Solve the convex problem $\mathcal{P}_1$ and obtain $\mathbf{b}^{(i+1)}$
            \item Set $i=i + 1$
        \end{itemize}
        
        \item \textbf{Until convergence:} $\| \mathbf{b}^{(i)} - \mathbf{b}^{(i-1)} \| < \epsilon$
        
        \item Compute objective:
        \[
            \mathcal{F}(s) = -s \Gamma^{\mathrm{th}} - \sum_{t=1}^T \log\left(1 - s \psi(t) \Omega_{\text{av}}\right) + \rho \widetilde{f}(\mathbf{b}^{(i)})
        \]
        
        \item \textbf{If} $\mathcal{F}(s) < \mathcal{F}_{\text{best}}$:
        \begin{itemize}
            \item Update: $\mathcal{F}_{\text{best}} \gets \mathcal{F}(s)$, $s^* \gets s$, $\mathbf{b}^* \gets \mathbf{b}^{(i)}$
        \end{itemize}
    \end{enumerate}

    \item \textbf{Output:} Optimal positions $\mathbf{b}^* = \{ \mathbf{b}^*(t) \}_{t=1}^T$ and optimal Chernoff parameter $s^*$
\end{enumerate}
\end{algorithm}

\section{Simulation Results} 
In this section, we present numerical simulations to evaluate the performance of the proposed PA position selection. We consider a square room with dimensions \(10 \times 10\) meters where the target and the communication user are located at $[6,-3,0]$ and $[2,2,0]$, respectively, and the DFRC-BS is positioned at $[0,0,d]$. The default simulation parameters are summarized in Table~\ref{tab:simulation_parameters}. We compare the proposed optimization-based scheme against two baseline methods. For baseline scheme 1, the PA is restricted to use the same position across all $T$ time slots. This position is optimized using \eqref{Convex_Problem} by enforcing identical position selection in each time slot. Baseline scheme 2 utilizes an antenna selection (AS) strategy, where the DFRC-BS is equipped with a transmit ULA consisting of $M$ fixed antennas spaced at $\lambda/2$, positioned at the center of the $x$-axis at $[5,0,d]$. For the proposed scheme, we evaluate the outage performance using both the closed-form expression derived in~\eqref{eq:outage_cdf} and Monte Carlo simulations based on the exact definition in \eqref{eq:outage1}. 

Fig.~2 illustrates the radar outage probability as a function of the transmit power for the proposed scheme and the two baseline methods. As expected, higher transmission power leads to reduced outage probability for all schemes, as stronger radar echoes improve target detectability. The proposed approach consistently achieves the lowest outage probability, thanks to its ability to activate PAs at different positions across time slots. This dynamic positioning enables the system to observe the target from multiple angular directions providing \textit{target diversity} which significantly improves robustness against low-RCS observation angles. A key trend in Fig.~2 is the performance gain achieved as the number of time slots $T$ increases. As $T$ increases from $2$ to $8$, the system acquires more independent angular views of the target, effectively improving the diversity order. This increase in diversity order enhances the system’s ability to withstand poor RCS condition for a single direction, leading to significantly improved sensing reliability.
Baseline scheme 1 employs a fixed PA position optimized for the average performance across all time slots. While it cannot exploit different look-angles, its selected location provides a more favorable pathloss to the target compared to baseline scheme 2. Baseline scheme 2 uses a fixed transmit ULA which results in a larger distance and higher pathloss for the target. As a result, baseline scheme 2 is outperformed by baseline scheme 1. Importantly, neither baseline scheme offers target diversity, as they provide the same look-angle in each time slot. In contrast, the proposed scheme dynamically changes the PA position across time slots, enabling angular diversity and significantly improved robustness to unfavorable RCS conditions. In addition, for $T=2$, we evaluate the closed-form outage expression derived in \eqref{eq:outage_cdf} using the PA positions obtained from the Chernoff bound-based optimization. The resulting outage probability closely aligns with the Monte Carlo simulation results obtained based on the exact outage definition in \eqref{eq:outage1}, confirming the accuracy of the analytical approximation.
\begin{figure}[t]\label{fig:outage_vs_threshold}
    \centering
\includegraphics[height= 6.15 cm, width=8.500cm]{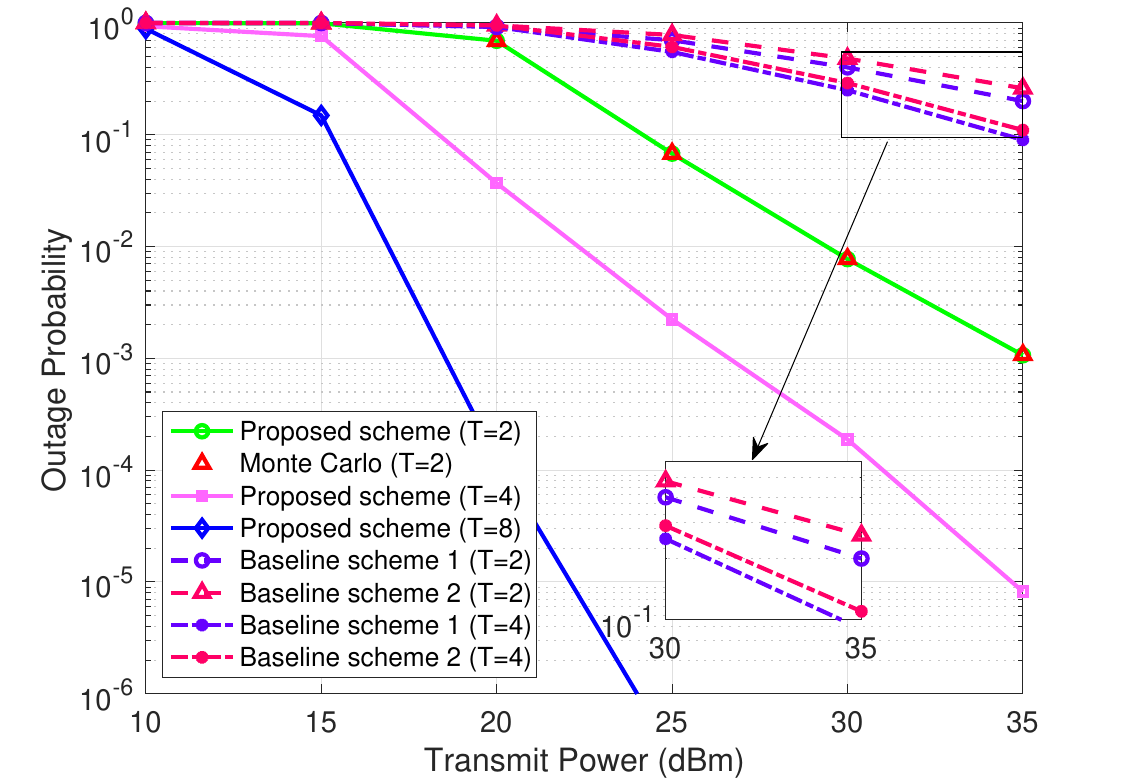}
\caption{Outage probability versus transmit power for $R_{\min}=0.5$~bps/Hz.}
\end{figure}
\begin{figure}[t]
    \centering
\includegraphics[width=8.500cm]{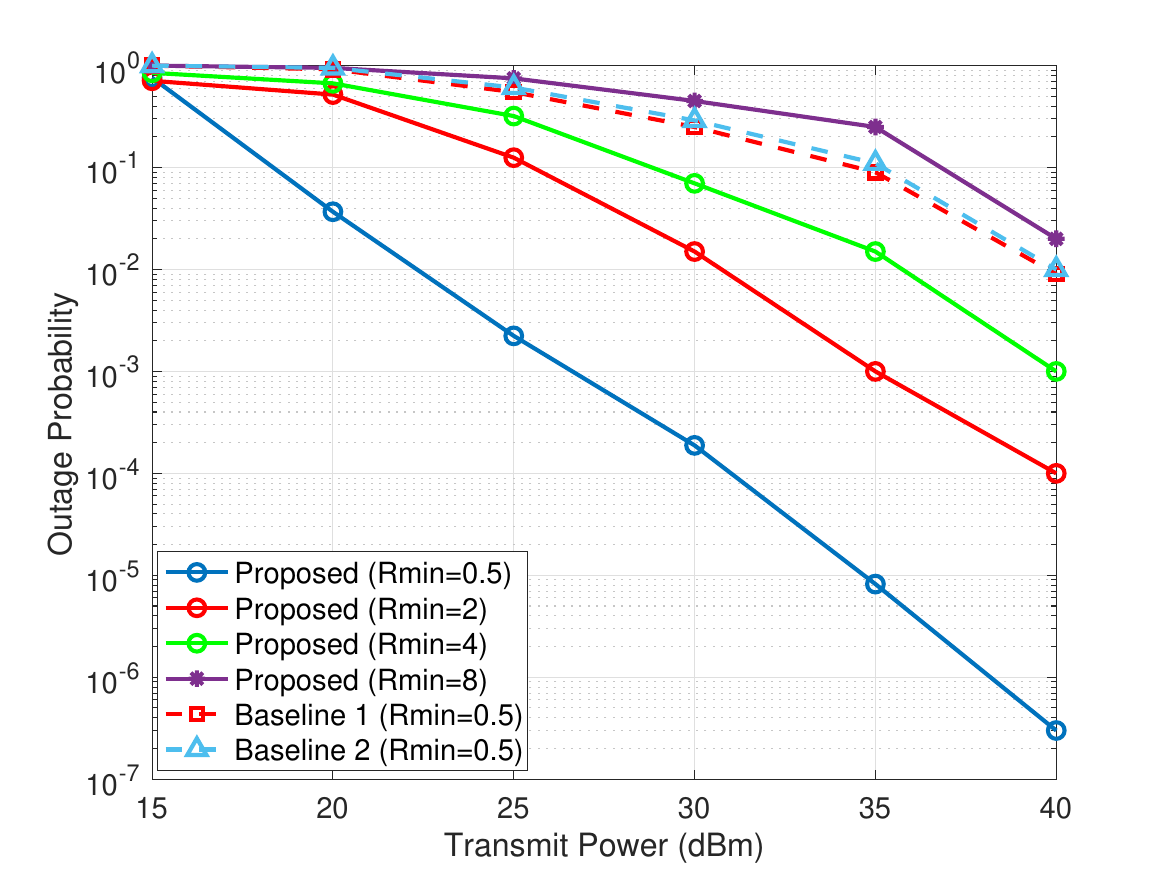}
\caption{Outage probability versus transmit power for $T=4$.}
\end{figure}

Fig.~3 illustrates the radar outage probability versus the transmit power for different minimum communication rate requirements $R_{\min}$. The proposed method consistently outperforms both baseline schemes, demonstrating the benefits of dynamically reconfigurable PA positions. However, a clear trade-off emerges as the required communication rate $R_{\min}$ increases. When $R_{\min}$ grows from $0.5$ to $2$ and $4$ bps/Hz, the outage performance of the proposed scheme gradually degrades. This is because higher communication QoS requirements force the system to prioritize PA positions with favorable communication channels, which may be located farther from the target. Although the number of angular observations and thus the target diversity remain unchanged due to the design of the system, the selected positions exhibit higher pathloss for sensing. This leads to weaker radar echoes and degraded sensing performance. For very strict communication QoS demands,  i.e., $R_{\min} = 8$ bps/Hz, this effect becomes more pronounced, as the optimization must prioritize communication even at the cost of sensing quality. This highlights a fundamental trade-off in ISAC system design: Ensuring strict communication reliability may require compromising sensing performance.

\begin{table}[t]
\caption{Simulation Parameters}
\label{tab:simulation_parameters}
\centering
\setlength{\tabcolsep}{5pt}
\begin{tabular}{|c|c|c|c|}
\hline
\textbf{Parameter} & \textbf{Value} & \textbf{Parameter} & \textbf{Value} \\
\hline
Waveguide length ($D_x$) & 10 m & Pinching positions ($M$) & 20 \\
Receive antennas ($N_R$) & 8 & SNR threshold ($\Gamma^{\text{th}}$) & 10 dB \\
Noise power ($\sigma^2$) & –90 dBm & RCS mean ($\Omega_{\mathrm{av}}$) & 1 m$^2$ \\
Carrier frequency ($f_c$) & 30 GHz & Attenuation ($\alpha$) & 0.18 \cite{docomo2019ntt} \\
Refractive index ($\eta_{\text{eff}}$) & 1.4 \cite{p3} & PA height ($d$) & 3 m \\
\hline
\end{tabular}
\end{table}

\section{Conclusion}
In this paper, we investigated, for the first time, PA-assisted ISAC systems with spatially discrete and dynamically reconfigurable antenna positions. To account for the inherent uncertainty introduced by RCS fluctuations, we proposed outage probability as a novel sensing performance metric for ISAC design. We formulated a joint optimization problem to minimize the radar outage probability by selecting different PA positions in different time slots, while ensuring the QoS of a communication user. Due to the non-convexity of the resulting problem, we developed a tractable solution based on a Chernoff bound surrogate and SCA. Simulation results show that the proposed approach outperforms two baseline schemes by leveraging temporal angular diversity, achieved through activating different PA positions over time. This enables multiple target views from different angles, improving sensing reliability. Our results also highlight a key trade-off: Stricter communication requirements limit flexibility in PA selection, reducing sensing performance. This underscores the value of jointly optimizing communication QoS and sensing reliability in PA-assisted ISAC systems. In future work, we will consider the impact of angular RCS correlations, moving beyond the current assumption of statistically independent RCSs across different observation angles.
\bibliographystyle{IEEEtran}
\bibliography{Ref.bib}

\begin{thebibliography}{10}
\providecommand{\url}[1]{#1}
\csname url@samestyle\endcsname
\providecommand{\newblock}{\relax}
\providecommand{\bibinfo}[2]{#2}
\providecommand{\BIBentrySTDinterwordspacing}{\spaceskip=0pt\relax}
\providecommand{\BIBentryALTinterwordstretchfactor}{4}
\providecommand{\BIBentryALTinterwordspacing}{\spaceskip=\fontdimen2\font plus
\BIBentryALTinterwordstretchfactor\fontdimen3\font minus
  \fontdimen4\font\relax}
\providecommand{\BIBforeignlanguage}[2]{{%
\expandafter\ifx\csname l@#1\endcsname\relax
\typeout{** WARNING: IEEEtran.bst: No hyphenation pattern has been}%
\typeout{** loaded for the language `#1'. Using the pattern for}%
\typeout{** the default language instead.}%
\else
\language=\csname l@#1\endcsname
\fi
#2}}
\providecommand{\BIBdecl}{\relax}
\BIBdecl

\bibitem{wong2020fluid}
K.-K. Wong, A.~Shojaeifard, K.-F. Tong, and Y.~Zhang, ``Fluid antenna
  systems,'' \emph{IEEE Trans. Wireless Commun.}, vol.~20, no.~3, pp.
  1950--1962, 2020.

\bibitem{zhu2023movable}
L.~Zhu, W.~Ma, B.~Ning, and R.~Zhang, ``Movable-antenna enhanced multiuser
  communication via antenna position optimization,'' \emph{IEEE Trans. Wireless
  Commun.}, vol.~23, no.~7, pp. 7214--7229, Jul. 2024.

\bibitem{khalili2024advanced}
A.~Khalili and R.~Schober, ``Advanced {ISAC} design: Movable antennas and
  accounting for dynamic {RCS},'' in \emph{Proc. IEEE GLOBECOM 2024}, 2024, pp.
  4022--4027.

\bibitem{docomo2019ntt}
H.~O.~Y. Suzuki and K.~Kawai, ``Pinching antenna: Using a dielectric waveguide
  as an antenna,'' \emph{NTT DOCOMO Technical J.}, 2022.

\bibitem{p1}
Z.~Ding, R.~Schober, and H.~Vincent~Poor, ``Flexible-antenna systems: A
  pinching-antenna perspective,'' \emph{IEEE Trans. Commun.}, pp. 1--1, 2025.

\bibitem{p3}
K.~Wang, Z.~Ding, and R.~Schober, ``Antenna activation for {NOMA} assisted
  pinching-antenna systems,'' \emph{IEEE Wireless Commun Letts.}, 2025.

\bibitem{p4}
S.~A. Tegos, P.~D. Diamantoulakis, Z.~Ding, and G.~K. Karagiannidis, ``Minimum
  data rate maximization for uplink pinching-antenna systems,'' \emph{IEEE
  Wireless Commun Letts.}, pp. 1--1, 2025.

\bibitem{Radar}
M.~I. Skolnik, ``Introduction to radar,'' \emph{Radar {H}andbook}, vol.~2,
  p.~21, 1962.

\bibitem{ISAC6G}
F.~Liu \emph{et~al.}, ``Integrated sensing and communications: Toward
  dual-functional wireless networks for {6G} and beyond,'' \emph{IEEE J.
  Select. Areas Commun.}, vol.~40, no.~6, pp. 1728--1767, Jun. 2022.

\bibitem{jsc-mimo-radar}
X.~Liu \emph{et~al.}, ``Joint transmit beamforming for multiuser {MIMO}
  communications and {MIMO} radar,'' \emph{IEEE Trans. Signal Process.},
  vol.~68, pp. 3929--3944, Jun. 2020.

\bibitem{mu-mimo-jsc}
F.~Liu \emph{et~al.}, ``{MU-MIMO} communications with {MIMO} radar: From
  co-existence to joint transmission,'' \emph{IEEE Trans. Wireless Commun.},
  vol.~17, no.~4, pp. 2755--2770, Apr. 2018.

\bibitem{Ang1}
A.~M. Haimovich, R.~S. Blum, and L.~J. Cimini, ``{MIMO} radar with widely
  separated antennas,'' \emph{IEEE Signal Processing Magazine}, vol.~25, no.~1,
  pp. 116--129, 2008.

\bibitem{Ang2}
E.~Fishler, A.~Haimovich, R.~Blum, R.~Cimini, D.~Chizhik, and R.~Valenzuela,
  ``Performance of {MIMO} radar systems: advantages of angular diversity,'' in
  \emph{Proc. IEEE Asilomar Conference on Signals, Systems and Computers},
  2004, pp. 305--309.

\bibitem{P_ISAC}
Y.~Qin, Y.~Fu, and H.~Zhang, ``Joint antenna position and transmit power
  optimization for pinching antenna-assisted {ISAC} systems,'' \emph{arXiv
  preprint arXiv:2503.12872}, 2025.

\bibitem{khalili2024efficient}
A.~Khalili, A.~Rezaei, D.~Xu, F.~Dressler, and R.~Schober, ``Efficient {UAV}
  hovering, resource allocation, and trajectory design for {ISAC} with limited
  backhaul capacity,'' \emph{IEEE Trans. Wireless Commun.}, vol.~23, no.~11,
  pp. 17\,635--17\,650, Nov. 2024.

\bibitem{akkouchi2008convolution}
M.~Akkouchi, ``On the convolution of exponential distributions,'' \emph{Journal
  of the Chungcheong Mathematical Society}, vol.~21, no.~4, pp. 501--501, 2008.

\end{thebibliography}
\end{document}